\documentclass[12pt]{article}

\begin{document}

\thispagestyle{empty}
\renewcommand{\thefootnote}{\fnsymbol{footnote}}

\begin{flushright}
{\small
SLAC--PUB--8358\\
January 2000\\}
\end{flushright}

\vspace{.8cm}

\begin{center}
{\bf\Large   
Formation of Patterns in Intense Hadron Beams. The Amplitude 
Equation Approach 
\footnote{Work supported by
Department of Energy contract  DE--AC03--76SF00515.}}

\vspace{1cm}

Stephan I. Tzenov\\
Stanford Linear Accelerator Center, Stanford University,
Stanford, CA  94309\\

\end{center}

\vfill

\begin{center}
{\bf\large Abstract }
\end{center}

\begin{quote}
We study the longitudinal motion of beam particles 
under the action of a single resonator wave induced 
by the beam itself. Based on the method of multiple 
scales we derive a system of coupled amplitude 
equations for the slowly varying part of the 
longitudinal distribution function and for the 
resonator wave envelope, corresponding to an 
arbitrary wave number. The equation governing the 
slow evolution of the voltage envelope is shown to 
be of Ginzburg--Landau type. 
\end{quote}

\vfill

\begin{center} 
{\it Paper presented at: } 

{\it Second ICFA Advanced Accelerator Workshop on } \\

{\it THE PHYSICS OF HIGH BRIGHTNESS BEAMS } \\

{\it UCLA Faculty Center, Los Angeles } \\

{\it November 9--12, 1999 } \\
\end{center}

\newpage

\pagestyle{plain}

\renewcommand{\theequation}{\thesection.\arabic{equation}}

\setcounter{equation}{0}

\section{Introduction}

So far, extensive work has been performed on the linear 
stability analysis of collective motion in particle accelerators 
\cite{chao}. Nonlinear theories \cite{az1}--\cite{cary} of wave 
interaction and formation of patterns and coherent structures 
in intense beams are however less prevalent, in part, due to the 
mathematical complexity of the subject, but also because of the 
commonly spread opinion that highly nonlinear regime is 
associated with poor machine performance that is best to be 
avoided.

Nevertheless, nonlinear wave interaction is a well observed 
phenomenon \cite{az1}, \cite{spentzouris} in present machines, 
complete and self-consistent theory explaining the processes, 
leading to the formation of self-organized structures mentioned 
above is far from being established. The present paper is aimed 
as an attempt in this direction.

The problem addressed here (perhaps, the simplest one) is the 
evolution of a beam in longitudinal direction under the 
influence of a resonator voltage induced by the beam itself. 
Linear theory is obviously unable to explain bunch (droplet) 
formation and bunch breakoff (especially in the highly damped 
regime), phenomena that have been observed by numerical 
simulations \cite{az1}, \cite{gerasimov}, \cite{cary}, but it 
should be considered as the first important step towards our 
final goal -- nonlinear model of wave interaction developed in 
Section 3. 

It is well-known that within the framework of linear stability 
analysis the solution of the original problem is represented 
as a superposition of plane waves with constant amplitudes, 
while the phases are determined by the spectrum of solutions 
to the dispersion equation. Moreover, the wave amplitudes are 
completely arbitrary and independent of the spatial and temporal 
variables. The effect of nonlinearities is to cause variation 
in the amplitudes in both space and time. We are interested 
in describing these variations, since they govern the relatively 
slow process of formation of self-organized patterns and 
coherent structures. 

The importance of the linear theory is embedded in the 
dispersion relation and the type of solutions it possesses. If 
the dispersion relation has no imaginary parts (no dissipation 
of energy occurs and no pumping from external energy sources 
is available) and its solutions, that is the wave frequency 
as a function of the wave number are all real, then the 
corresponding amplitude equations describing the evolution 
of the wave envelopes will be of nonlinear Schr\"{o}dinger 
type. Another possibility arises for conservative systems when 
some of the roots of the dispersion equation appear in complex 
conjugate pairs. Then the amplitude equations can be shown to 
be of the so called AB--type \cite{dodd}. For open systems 
(like the system studied here) the dispersion relation is in 
general a complex valued function of the wave frequency and 
wave number and therefore its solutions will be complex. It 
can be shown \cite{dodd} that the equation governing the slow 
evolution of the wave amplitudes in this case will be the 
Ginzburg--Landau equation.

Based on the renormalization group approach we have recently 
derived a Ginzburg--Landau equation for the amplitude of the 
resonator voltage in the case of a coasting beam \cite{az3}. 
The derivation has been carried out under the assumption that 
the spatial evolution of the system is much slower compared to 
the temporal one. This restriction has been removed here, and 
the present paper may be considered as an extension of \cite{az3}. 

Using the method of multiple scales we derive a set of coupled 
amplitude equations for the slowly varying part of the 
longitudinal distribution function and for the intensity of a 
single resonator wave with an arbitrary wave number (and wave 
frequency, specified as a solution to the linear dispersion 
equation). The equation governing the evolution of the voltage 
envelope is shown to be of Ginzburg--Landau type. 

\setcounter{equation}{0}

\section{Formulation of the Problem}

It is well-known that the longitudinal dynamics of an 
individual beam particle is governed by the set of equations 
\cite{bruck} 

\begin{eqnarray}
{\frac{dz_1} {dt}} = k_0 \Delta E \qquad ; \qquad 
{\frac{d \Delta E} {dt}} = {\frac{e \omega_s V_{RF}} {2 \pi}} 
{\left( \sin \phi - \sin \phi_s \right)} + 
{\frac{e \omega_s} {2 \pi}} V_1, 
\label{Hamilton} 
\end{eqnarray}

\noindent
where

\begin{eqnarray}
k_0 = -{\frac{\eta \omega_s} {\beta_s^2 E_s}}
\end{eqnarray}

\noindent
is the proportionality constant between the frequency deviation 
of a non synchronous particle with respect to the frequency 
$\omega_s$ of the synchronous one, and the energy deviation 
$\Delta E = E - E_s$. The quantity $k_0$ also involves the 
phase slip coefficient 
$\eta = \alpha_M - \gamma_s^{-2}$, where $\alpha_M$ is the 
momentum compaction factor \cite{bruck}. The variables

\begin{eqnarray}
z_1 = \theta - \omega_s t  \qquad ; \qquad 
\phi = \phi_s - h z_1. 
\label{Longpos} 
\end{eqnarray}

\noindent
are the azimuthal displacement of the particle with respect to 
the synchronous one, and the phase of the RF field, respectively. 
Here $V_{RF}$ is the amplitude of the RF voltage and $h$ is the 
harmonic number. Apart from the RF field we assume that beam 
motion is influenced by a resonator voltage $V_1$ due to a broad 
band impedance 

\begin{eqnarray}
{\frac {\partial^2 V_1} {\partial z_1^2}} - 
2 \gamma {\frac {\partial V_1} {\partial z_1}} + 
\omega^2 V_1 = 
{\frac {2 \gamma e {\cal R}} {\omega_s}} 
{\frac {\partial I_1} {\partial t}}, 
\label{Imped} 
\end{eqnarray}

\noindent
where

\begin{eqnarray}
\omega = {\frac {\omega_r} {\omega_s}}  \qquad ; \qquad 
\gamma = {\frac {\omega} {2 Q}}  \qquad ; \qquad 
I_1{\left( \theta; t \right)} = 
\int d \Delta E {\left( \omega_s + k_0 \Delta E \right)} 
f_1{\left( \theta, \Delta E; t \right)}, 
\label{Const} 
\end{eqnarray}

\noindent
$f_1{\left( \theta, \Delta E; t \right)}$ is the longitudinal 
distribution function, $\omega_r$ is the resonant frequency, 
$Q$ is the quality factor of the resonator and ${\cal R}$ is 
the resonator shunt impedance. 

It is convenient to pass to a new independent variable (``time'') 
$\theta$ and to the new dimensionless variables \cite{az1}, 
\cite{az4}:

\begin{eqnarray}
\tau = \nu_s \theta  \qquad ; \qquad 
z = z_1 {\sqrt{\nu_s}}  \qquad ; \qquad 
u = {\frac {1} {\sqrt{\nu_s}}} {\frac {k_0 \Delta E} 
{\omega_s}}, 
\label{Dimless1} 
\end{eqnarray}

\begin{eqnarray}
f_1{\left( \theta, \Delta E; t \right)} = 
{\frac {\rho_0 {\left| k_0 \right|}} 
{\omega_s {\sqrt{\nu_s}}}}
f{\left( z, u; \theta \right)}  \qquad ; \qquad 
V_1 = \lambda_1 V  \qquad ; \qquad 
I_1 = \omega_s \rho_0 I, 
\label{Dimless2} 
\end{eqnarray}

\noindent
where

\begin{eqnarray}
\nu_s^2 = {\frac {eh k_0 V_{RF} \cos \phi_s} 
{2 \pi \omega_s}}  \qquad ; \qquad 
\lambda_1 = {2 \gamma_0 e {\cal R} \omega_s \rho_0}. 
\label{Nuslamb1} 
\end{eqnarray}

\noindent
In the above expressions the quantity $\rho_0$ is the uniform 
beam density in the thermodynamic limit. The linearized 
equations of motion (\ref{Hamilton}) and equation (\ref{Imped}) 
in these variables read as:

\begin{eqnarray}
{\frac {dz} {d \tau}} = u  \qquad ; \qquad 
{\frac {du} {d \tau}} = -z + \lambda V, 
\label{Motion} 
\end{eqnarray}

\begin{eqnarray}
{\frac {\partial^2 V} {\partial z^2}} - 
2 \gamma_0 {\frac {\partial V} {\partial z}} + 
\omega_0^2 V = - 
{\frac {\partial I} {\partial z}}  \qquad ; \qquad 
I{\left( z; \theta \right)} = 
\int du {\left( 1 + u {\sqrt{\nu_s}} \right)} 
f{\left( z, u; \theta \right)}, 
\label{Impedance1} 
\end{eqnarray}

\noindent
where

\begin{eqnarray}
\gamma_0 = {\frac {\gamma} {\sqrt{\nu_s}}}  \qquad ; \qquad 
\omega_0 = {\frac {\omega} {\sqrt{\nu_s}}}  \qquad ; \qquad 
\lambda = {\frac {e^2 {\cal R} \gamma_0 k_0 \rho_0} 
{\pi \nu_s {\sqrt{\nu_s}}}}. 
\label{Scaleconst} 
\end{eqnarray}

\noindent
We can now write the Vlasov equation for the longitudinal 
distribution function $f{\left( z, u; \theta \right)}$, which 
combined with the equation for the resonator voltage 
$V{\left( z; \theta \right)}$

\begin{eqnarray}
{\frac {\partial f} {\partial \tau}} + 
u {\frac {\partial f} {\partial z}} - 
z {\frac {\partial f} {\partial u}} + 
\lambda V {\frac {\partial f} {\partial u}} = 0, 
\label{Vlasov} 
\end{eqnarray}

\begin{eqnarray}
{\frac {\partial^2 V} {\partial z^2}} - 
2 \gamma_0 {\frac {\partial V} {\partial z}} + 
\omega_0^2 V = - 
{\frac {\partial I} {\partial z}}, 
\label{Impedance}
\end{eqnarray}

\begin{eqnarray}
I{\left( z; \theta \right)} = 
\int du {\left( 1 + u {\sqrt{\nu_s}} \right)} 
f{\left( z, u; \theta \right)}, 
\label{Current} 
\end{eqnarray}

\noindent
comprises the starting point for our subsequent analysis.

\setcounter{equation}{0}

\section{Derivation of the Amplitude Equations for a Coasting 
Beam}

In this Section we analyze the simplest case of a coasting beam. 
The model equations (\ref{Vlasov}) and (\ref{Impedance}) acquire 
the form \cite{az4} 

\begin{eqnarray}
{\frac {\partial f} {\partial \theta}} + 
u {\frac {\partial f} {\partial z}} + 
\lambda V {\frac {\partial f} {\partial u}} = 0, 
\label{Vlascoast} 
\end{eqnarray}

\begin{eqnarray}
{\frac {\partial^2 V} {\partial z^2}} - 
2 \gamma {\frac {\partial V} {\partial z}} + 
\omega^2 V = 
\int du {\left( 
{\frac {\partial f} {\partial \theta}} - 
{\frac {\partial f} {\partial z}} \right)}, 
\label{Impedcoast}
\end{eqnarray}

\noindent
where the parameter $\lambda$ should be calculated for 
$\nu_s = 1$. In what follows it will be convenient to write the 
above equations more compactly as: 

\begin{eqnarray}
{\widehat{\cal F}}{\left( {\frac {\partial} {\partial \theta}}, 
~{\frac {\partial} {\partial z}}, ~u  \right)} f + 
\lambda V {\frac {\partial f} {\partial u}} = 0, 
\label{Mvlasov} 
\end{eqnarray}

\begin{eqnarray}
{\widehat{\cal V}}{\left( {\frac {\partial} {\partial z}}, 
~\omega \right)} V = 
{\widehat{\cal L}}{\left( {\frac {\partial} {\partial \theta}}, 
~{\frac {\partial} {\partial z}} \right)} 
{\left \langle f \right \rangle}, 
\label{Mimpedance}
\end{eqnarray}

\noindent
where we have introduced the linear operators

\begin{eqnarray}
{\widehat{\cal F}}{\left( {\frac {\partial} {\partial \theta}}, 
~{\frac {\partial} {\partial z}}, ~u  \right)} = 
{\frac {\partial} {\partial \theta}} + 
u {\frac {\partial} {\partial z}}, 
\label{OF} 
\end{eqnarray}

\begin{eqnarray}
{\widehat{\cal V}}{\left( {\frac {\partial} {\partial z}}, 
~\omega \right)} = 
{\frac {\partial^2} {\partial z^2}} - 
2 \gamma {\frac {\partial} {\partial z}} + \omega^2, 
\label{OV} 
\end{eqnarray}

\begin{eqnarray}
{\widehat{\cal L}}{\left( {\frac {\partial} {\partial \theta}}, 
~{\frac {\partial} {\partial z}} \right)} = 
{\frac {\partial} {\partial \theta}} - 
{\frac {\partial} {\partial z}}, 
\label{OL} 
\end{eqnarray}

\begin{eqnarray}
{\left \langle {\cal G} {\left( z, u; \theta \right)} 
\right \rangle} = 
\int du {\cal G} {\left( z, u; \theta \right)}. 
\label{OBracket} 
\end{eqnarray}

\noindent
To obtain the desired amplitude equation for nonlinear waves 
we use the method of multiple scales \cite{dodd}, \cite{debnath}. 
The key point of this approach is to introduce slow temporal 
as well as spatial scales according to the relations:

\begin{eqnarray}
\theta  \quad ; \quad 
T_1 = \epsilon \theta  \quad ; \quad 
T_2 = \epsilon^2 \theta  \quad ; \quad
\ldots  \quad ; \quad 
T_n = \epsilon^n \theta  \quad ; \quad 
\ldots 
\label{Scaletheta} 
\end{eqnarray}

\begin{eqnarray}
z  \quad ; \quad 
z_1 = \epsilon z  \quad ; \quad 
z_2 = \epsilon^2 z  \quad ; \quad 
\ldots  \quad ; \quad 
z_n = \epsilon^n z  \quad ; \quad 
\ldots 
\label{Scalez} 
\end{eqnarray}

\noindent
where $\epsilon$ is a formal small parameter. Next is to utilize 
the perturbation expansion of the longitudinal distribution 
function $f$, the resonator voltage $V$ 

\begin{eqnarray}
f = f_0{\left( u \right)} + \sum \limits_{k=1}^{\infty} 
\epsilon^k f_k  \qquad ; \qquad 
V = \sum \limits_{k=1}^{\infty} \epsilon^k V_k, 
\label{ExpandfV} 
\end{eqnarray}

\noindent
and the operator expansions

$$
{\widehat{\cal F}}{\left( 
{\frac {\partial} {\partial \theta}} + 
\sum \limits_{k=1}^{\infty} \epsilon^k 
{\frac {\partial} {\partial T_k}}, ~
{\frac {\partial} {\partial z}} + 
\sum \limits_{k=1}^{\infty} \epsilon^k 
{\frac {\partial} {\partial z_k}}, ~u \right)} = 
$$

\begin{eqnarray}
= {\widehat{\cal F}}{\left( 
{\frac {\partial} {\partial \theta}}, ~
{\frac {\partial} {\partial z}}, ~u \right)} + 
\sum \limits_{k=1}^{\infty} \epsilon^k 
{\widehat{\cal F}}{\left( 
{\frac {\partial} {\partial T_k}}, ~
{\frac {\partial} {\partial z_k}}, ~u \right)}, 
\label{ExpandF} 
\end{eqnarray}

$$
{\widehat{\cal L}}{\left( 
{\frac {\partial} {\partial \theta}} + 
\sum \limits_{k=1}^{\infty} \epsilon^k 
{\frac {\partial} {\partial T_k}}, ~
{\frac {\partial} {\partial z}} + 
\sum \limits_{k=1}^{\infty} \epsilon^k 
{\frac {\partial} {\partial z_k}} \right)} = 
$$

\begin{eqnarray}
= {\widehat{\cal L}}{\left( 
{\frac {\partial} {\partial \theta}}, ~
{\frac {\partial} {\partial z}} \right)} + 
\sum \limits_{k=1}^{\infty} \epsilon^k 
{\widehat{\cal L}}{\left( 
{\frac {\partial} {\partial T_k}}, ~
{\frac {\partial} {\partial z_k}} \right)}, 
\label{ExpandL} 
\end{eqnarray}

\begin{eqnarray}
{\widehat{\cal V}}{\left( 
{\frac {\partial} {\partial z}} + 
\sum \limits_{k=1}^{\infty} \epsilon^k 
{\frac {\partial} {\partial z_k}} \right)} = 
{\widehat{\cal V}} + \epsilon {\widehat{\cal V}}_z 
{\frac {\partial} {\partial z_1}} + 
{\frac {\epsilon^2} {2}} {\left( 
{\widehat{\cal V}}_{zz} 
{\frac {\partial^2} {\partial z_1^2}} + 
2 {\widehat{\cal V}}_z {\frac {\partial} {\partial z_2}} 
\right)} + \ldots 
\label{ExpandV} 
\end{eqnarray}

\noindent
where ${\widehat{\cal V}}_z$ implies differentiation with 
respect to $\partial / \partial z$. Substituting them back 
into (\ref{Mvlasov}) and (\ref{Mimpedance}) we obtain the 
corresponding perturbation equations order by order. It is 
worth noting that without loss of generality we can miss 
out the spatial scale $z_2$, because it can be transformed 
away by a simple change of the reference frame. For the sake 
of saving space we will omit the explicit substitution and 
subsequent calculations and state the final result order by 
order.

\underline{First order $O(\epsilon)$:}

\begin{eqnarray}
{\widehat{\cal F}} f_1 + \lambda V_1 {\frac 
{\partial f_0} {\partial u}} = 0, 
\label{Order11} 
\end{eqnarray}

\begin{eqnarray}
{\widehat{\cal V}} V_1 = 
{\widehat{\cal L}} 
{\left \langle f_1 \right \rangle}. 
\label{Order12} 
\end{eqnarray}

\underline{Second order $O(\epsilon^2)$:}

\begin{eqnarray}
{\widehat{\cal F}} f_2 + \lambda V_2 {\frac 
{\partial f_0} {\partial u}} = -{\widehat{\cal F}}_1 f_1 - 
\lambda V_1 {\frac {\partial f_1} {\partial u}}, 
\label{Order21} 
\end{eqnarray}

\begin{eqnarray}
{\widehat{\cal V}} V_2 = 
{\widehat{\cal L}} 
{\left \langle f_2 \right \rangle} + 
{\widehat{\cal L}}_1 
{\left \langle f_1 \right \rangle} - {\widehat{\cal V}}_z 
{\frac {\partial V_1} {\partial z_1}}. 
\label{Order22} 
\end{eqnarray}

\underline{Third order $O(\epsilon^3)$:}

\begin{eqnarray}
{\widehat{\cal F}} f_3 + \lambda V_3 {\frac 
{\partial f_0} {\partial u}} = -{\widehat{\cal F}}_1 f_2 - 
{\widehat{\cal F}}_2 f_1 - 
\lambda V_1 {\frac {\partial f_2} {\partial u}} - 
\lambda V_2 {\frac {\partial f_1} {\partial u}}, 
\label{Order31} 
\end{eqnarray}

\begin{eqnarray}
{\widehat{\cal V}} V_3 = 
{\widehat{\cal L}} 
{\left \langle f_3 \right \rangle} + 
{\widehat{\cal L}}_1 
{\left \langle f_2 \right \rangle} + 
{\widehat{\cal L}}_2 
{\left \langle f_1 \right \rangle} - {\widehat{\cal V}}_z 
{\frac {\partial V_2} {\partial z_1}} - 
{\frac {{\widehat{\cal V}}_{zz}} {2}} 
{\frac {\partial^2 V_1} {\partial z_1^2}}, 
\label{Order32}
\end{eqnarray}

\noindent
where ${\widehat{\cal F}}_n$ and ${\widehat{\cal L}}_n$ 
are the corresponding operators, calculated for 
$T_n$ and $z_n$. 

In order to solve consistently the perturbation equations for 
each order we need a unique equation for one of the unknowns; 
it is more convenient to have a sole equation for the 
distribution functions $f_n$ alone. This will prove later to 
be very efficient for the removal of secular terms that 
appear in higher orders. By inspecting the above equations 
order by order one can catch their general form: 

\begin{eqnarray}
{\widehat{\cal F}} f_n + \lambda V_n {\frac 
{\partial f_0} {\partial u}} = \alpha_n  \qquad ; \qquad 
{\widehat{\cal V}} V_n = 
{\widehat{\cal L}} 
{\left \langle f_n \right \rangle} + \beta_n, 
\label{GenForm} 
\end{eqnarray}

\noindent
where $\alpha_n$ and $\beta_n$ are known functions, 
determined from previous orders. Eliminating $V_n$ we 
obtain:

\begin{eqnarray}
{\widehat{\cal V}}{\widehat{\cal F}} f_n + 
\lambda {\frac {\partial f_0} {\partial u}} 
{\widehat{\cal L}} 
{\left \langle f_n \right \rangle} = - \lambda 
{\frac {\partial f_0} {\partial u}} \beta_n + 
{\widehat{\cal V}} \alpha_n. 
\label{Fnequation} 
\end{eqnarray}

\noindent
Let us now proceed with solving the perturbation equations. 
The analysis of the first order equations (linearized 
equations) is quite standard, and for the one-wave solution 
we readily obtain:

\begin{eqnarray}
V_1 = E{\left( z_n; T_n \right)} e^{i \varphi} + 
E^{\ast}{\left( z_n; T_n \right)} e^{-i \varphi^{\ast}}, 
\label{Sol1V} 
\end{eqnarray}

\begin{eqnarray}
f_1 = -\lambda {\frac {\partial f_0} {\partial u}} 
{\left[ {\frac {E{\left( z_n; T_n \right)}} 
{{\widetilde{\cal F}} {\left( i \Omega, -ik, u \right)}}} 
e^{i \varphi} + 
{\frac {E^{\ast}{\left( z_n; T_n \right)}} 
{{\widetilde{\cal F}}^{\ast} {\left( i \Omega, -ik, u \right)}}} 
e^{-i \varphi^{\ast}} \right]} + 
F{\left( z_n, u; T_n \right)}, 
\label{Sol1f} 
\end{eqnarray}

\noindent
with 

\begin{eqnarray}
\varphi = \Omega \theta - kz, 
\label{Phase} 
\end{eqnarray}

\noindent
where given the wave number $k$, the wave frequency 
$\Omega(k)$ is a solution to the dispersion equation: 

\begin{eqnarray}
{\widetilde{\cal D}} {\left( k, \Omega(k) \right)} 
\equiv 0.
\label{Dispeq} 
\end{eqnarray}

\noindent
The dispersion function 
${\widetilde{\cal D}} {\left( k, \Omega \right)}$ is 
proportional to the dielectric permittivity of the beam 
and is given by the expression

\begin{eqnarray}
{\widetilde{\cal D}} {\left( k, \Omega \right)} = 
{\widetilde{\cal V}} {\left( -ik \right)} + \lambda 
{\widetilde{\cal L}} {\left( i \Omega, -ik \right)} 
{\left \langle {\frac {1} 
{{\widetilde{\cal F}} {\left( i \Omega, -ik, u \right)}}} 
{\frac {\partial f_0} {\partial u}} \right \rangle}, 
\label{Dispfunc} 
\end{eqnarray}

\noindent
where

\begin{eqnarray}
{\widehat{\cal F}} e^{i \varphi} = 
{\widetilde{\cal F}} {\left( i \Omega, -ik, u \right)} 
e^{i \varphi}  \quad ; \quad 
{\widehat{\cal V}} e^{i \varphi} = 
{\widetilde{\cal V}} {\left( -ik \right)} e^{i \varphi}  
\quad ; \quad 
{\widehat{\cal L}} e^{i \varphi} = 
{\widetilde{\cal L}} {\left( i \Omega, -ik \right)} 
e^{i \varphi}. 
\label{FVLtilde} 
\end{eqnarray}

\noindent
Note that the wave frequency has the following symmetry 
property: 

\begin{eqnarray}
\Omega^{\ast}(k) = - \Omega(-k).
\label{Frequency} 
\end{eqnarray}

\noindent
The functions $E{\left( z_n; T_n \right)}$ and 
$F{\left( z_n, u; T_n \right)}$  in equations (\ref{Sol1V}) 
and (\ref{Sol1f}) are the amplitude function we wish to 
determine. Clearly, these functions are constants with 
respect to the fast scales, but to this end they are allowed 
to be generic functions of the slow ones.

In order to specify the dependence of the amplitude functions 
on the slow scales, that is to derive the desired amplitude 
equations one need to go beyond the first order. The first 
step is to evaluate the right hand side of equation 
(\ref{Fnequation}) corresponding to the second order with 
the already found solution (\ref{Sol1V}) and (\ref{Sol1f}) 
for the first order. This yields terms (proportional to 
$e^{i \varphi}$) belonging to the kernel of the linear 
operator on the left hand side of equation (\ref{Fnequation}), 
which consequently give rise to the so called secular 
contributions to the perturbative solution. If the spectrum 
of solutions to the dispersion equation (\ref{Dispeq}) is 
complex (as is in our case), terms proportional to 
$e^{-2 Im(\Omega) \theta}$ appear on the right hand side 
of (\ref{Fnequation}). Since, the imaginary part of the wave 
frequency we consider small, the factor 
$e^{-2 Im(\Omega) \theta}$ is slowly varying in $\theta$ and we 
can replace it by $e^{-2 Im(\Omega) T_n}$, where the slow 
temporal scale $T_n$ is to be specified later. This in turn 
produces additional secular terms, which need to be taken 
care of as well. (Note that exactly for this purpose we have 
chosen two amplitude functions at first order). The procedure 
to avoid secular terms is to impose certain conditions on the 
amplitudes 
$E{\left( z_n; T_n \right)}$ and $F{\left( z_n, u; T_n \right)}$, 
that guarantee exact cancellation of all terms proportional 
to $e^{i \varphi}$ and terms constant in the fast scales $z$ 
and $\theta$ (containing $e^{-2 Im(\Omega) T_n}$) on the right 
hand side of equation (\ref{Fnequation}). One can easily check 
by direct calculation that the above mentioned conditions 
read as: 

\begin{eqnarray}
{\frac {\partial {\widetilde{\cal D}}} {\partial \Omega}} 
{\frac {\partial E} {\partial T_1}} - 
{\frac {\partial {\widetilde{\cal D}}} {\partial k}} 
{\frac {\partial E} {\partial z_1}} = - i \lambda 
{\widetilde{\cal L}} 
{\left \langle {\frac {1} {{\widetilde{\cal F}}}} 
{\frac {\partial F} {\partial u}} \right \rangle} E, 
\label{Symmetry1} 
\end{eqnarray}

\begin{eqnarray}
{\widehat{\cal F}}_1 F + 2 \lambda^2 Im(\Omega) 
{\frac {\partial} {\partial u}} 
{\left( 
{\frac {1} 
{{\left | {\widetilde{\cal F}} \right |}^2}} 
{\frac {\partial f_0} {\partial u}} \right)} 
{\left | E \right |}^2 
e^{-2 Im(\Omega) T_n} = - {\frac {\lambda} {\omega^2}} 
{\frac {\partial f_0} {\partial u}} 
{\widehat{\cal L}}_1 
{\left \langle F \right \rangle}. 
\label{Symmetry2} 
\end{eqnarray}

\noindent
Noting that the group velocity of the wave 
$\Omega_g = d \Omega / dk$ is given by 

\begin{eqnarray}
{\frac {\partial {\widetilde{\cal D}}} {\partial k}} + 
{\frac {\partial {\widetilde{\cal D}}} {\partial \Omega}} 
{\frac {d \Omega} {dk}} = 0  \qquad \Longrightarrow \qquad 
\Omega_g = - 
{\frac {\partial {\widetilde{\cal D}}} {\partial k}} 
{\left( 
{\frac {\partial {\widetilde{\cal D}}} {\partial \Omega}}
\right)}^{-1} 
\label{Groupvel} 
\end{eqnarray}

\noindent
we get

\begin{eqnarray}
{\frac {\partial E} {\partial T_1}} + 
\Omega_g {\frac {\partial E} {\partial z_1}} = - i \lambda 
{\left( 
{\frac {\partial {\widetilde{\cal D}}} {\partial \Omega}}
\right)}^{-1} {\widetilde{\cal L}} 
{\left \langle {\frac {1} {{\widetilde{\cal F}}}} 
{\frac {\partial F} {\partial u}} \right \rangle} E. 
\label{Symmetry} 
\end{eqnarray}

\noindent
The above equations (\ref{Symmetry2}) and (\ref{Symmetry}) 
are the amplitude equations to first order. Note that if 
$Im(\Omega) = 0$ we could simply set $F$ equal to zero and 
then equation (\ref{Symmetry}) would describe the symmetry 
properties of the original system (\ref{Vlascoast}) and 
(\ref{Impedcoast}) with respect to a linear plane wave 
solution. However, we are interested in the nonlinear 
interaction between waves (of increasing harmonicity) 
generated order by order, and as it can be easily seen 
the first nontrivial result taking into account this 
interaction will come out at third order. To pursue 
this we need the explicit (non secular) second order 
solutions for $f_2$ and $V_2$. 

Solving the second order equation (\ref{Fnequation}) with 
the remaining non secular part of the second order right 
hand side and then solving equation (\ref{Order22}) with 
the already determined $f_2$ we find

\begin{eqnarray}
f_2 = S_F{\left( k, \Omega, u \right)} E^2 
e^{2i \varphi} + c.c. + 
F_2{\left( z_n, u; T_n \right)}, 
\label{Sol2f} 
\end{eqnarray}

\begin{eqnarray}
V_2 = S_V{\left( k, \Omega \right)} E^2 
e^{2i \varphi} + f_V e^{i \varphi} + c.c. + 
G_V {\left( z_n, T_n; {\left[ F \right]} \right)}, 
\label{Sol2V} 
\end{eqnarray}

\noindent
where $c.c.$ denotes complex conjugation. Without loss of 
generality we can set the generic function 
$F_2{\left( z_n, u; T_n \right)}$ equal to zero. Note that, 
in case $Im(\Omega) = 0$ we could have set $F = 0$, as 
mentioned earlier, but we should keep the function $F_2$ nonzero 
in order to cancel third order secular terms depending on the 
slow scales only. Moreover, the functions $S_F$, $S_V$, $f_V$ 
and the functional $G_V{\left( {\left[ F \right]} \right)}$ 
of the amplitude $F$ are given by the following expressions: 

\begin{eqnarray}
S_F{\left( k, \Omega, u \right)} = {\frac {\lambda^2} {2}} 
{\frac {{\widetilde{\cal V}}(-2ik)} 
{{\widetilde{\cal D}}(2k, 2 \Omega)}} 
{\frac {1} {{\widetilde{\cal F}}(i \Omega, -ik, u)}} 
{\frac {\partial} {\partial u}} {\left[ 
{\frac {1} {{\widetilde{\cal F}}(i \Omega, -ik, u)}} 
{\frac {\partial f_0} {\partial u}} \right]}, 
\label{SF} 
\end{eqnarray}

\begin{eqnarray}
S_V{\left( k, \Omega \right)} = \lambda^2 
{\frac {{\widetilde{\cal L}} {\left( i \Omega, -ik \right)}} 
{{\widetilde{\cal D}} {\left( 2k, 2 \Omega \right)}}} 
{\left \langle 
{\frac {1} {{\widetilde{\cal F}}(i \Omega, -ik, u)}} 
{\frac {\partial} {\partial u}} {\left[ 
{\frac {1} {{\widetilde{\cal F}}(i \Omega, -ik, u)}} 
{\frac {\partial f_0} {\partial u}} \right]} 
\right \rangle}, 
\label{SV} 
\end{eqnarray}

\begin{eqnarray}
f_V = {\frac {i} {{\widetilde{\cal V}}(-ik)}} 
{\left[ i \lambda {\widetilde{\cal I}} 
{\widehat{\cal L}}_1 E - 
{\widetilde{\cal V}}_k(-ik) 
{\frac {\partial E} {\partial z_1}} \right]}, 
\label{FV} 
\end{eqnarray}

\begin{eqnarray}
G_V {\left( z_n, T_n; {\left[ F \right]} \right)} = 
{\frac {1} {\omega^2}} {\widehat{\cal L}}_1 
{\left \langle F \right \rangle}, 
\label{GV} 
\end{eqnarray}

\begin{eqnarray}
{\widetilde{\cal I}}(k, \Omega) = 
{\left \langle 
{\frac {1} {{\widetilde{\cal F}}(i \Omega, -ik, u)}} 
{\frac {\partial f_0} {\partial u}} \right \rangle}, 
\label{Itilde} 
\end{eqnarray}

\noindent
where the $k$-index implies differentiation with respect to 
$k$. 

The last step consists in evaluating the right hand side of 
equation (\ref{Fnequation}), corresponding to the third 
order with the already found first and second order solutions. 
Removal of secular terms in the slow scales leads us finally 
to the amplitude equation for the function 
$F{\left( z_n, u; T_n \right)}$, that is

$$
{\frac {\partial} {\partial T_2}} 
{\left( \omega^2 F + \lambda 
{\left \langle F \right \rangle} 
{\frac {\partial f_0} {\partial u}} 
\right)} + {\frac {2 \lambda \gamma} {\omega^2}} 
{\frac {\partial f_0} {\partial u}} 
{\frac {\partial} {\partial z_1}} 
{\widehat{\cal L}}_1 {\left \langle F \right \rangle} + 
\lambda {\frac {\partial F} {\partial u}}
{\widehat{\cal L}}_1 {\left \langle F \right \rangle} = 
$$

\begin{eqnarray}
= \lambda^2 \omega^2 
{\left[ {\frac {\partial} {\partial u}} 
{\left( {\frac {1} {{\widetilde{\cal F}}^{\ast}}} 
{\frac {\partial f_0} {\partial u}} 
\right)} f_V E^{\ast} + 
{\frac {\partial} {\partial u}} 
{\left( {\frac {1} {{\widetilde{\cal F}}}} 
{\frac {\partial f_0} {\partial u}} 
\right)} f_V^{\ast} E 
\right]} e^{-2 Im(\Omega) T_2}. 
\label{Amplitude1} 
\end{eqnarray}

\noindent
Elimination of secular terms in the fast scales leads us to 
a generalized cubic Ginzburg--Landau type of equation for the 
amplitude $E{\left( z_n, T_n \right)}$:

$$
i {\frac {\partial {\widetilde{\cal D}}} {\partial \Omega}} 
{\frac {\partial E} {\partial T_2}} = {\cal A} 
{\frac {\partial^2 E} {\partial z_1^2}} + \lambda a 
{\frac {\partial} {\partial z_1}} 
{\left \{ {\cal G}{\left( {\left[ F \right]} \right)}E 
\right \}} + \lambda {\cal B} {\left | E \right |}^2 E 
e^{-2 Im(\Omega) T_2} - 
$$

\begin{eqnarray}
- \lambda^2 {\cal C} G_V {\left( 
{\left[ F \right]} \right)} E + \lambda 
{\widetilde{\cal L}} 
{\cal G}{\left( {\left[ F \right]} \right)} f_V, 
\label{Amplitude2} 
\end{eqnarray}

\noindent
where the coefficients $a(k)$, ${\cal A}(k)$, ${\cal B}(k)$ 
and ${\cal C}(k)$ are given by the expressions:

\begin{eqnarray}
a(k) = {\widetilde{\cal V}}_k 
{\left(
{\frac {\partial {\widetilde{\cal D}}} {\partial \Omega}}
\right)}^{-1}, 
\label{Ascoeff} 
\end{eqnarray}

\begin{eqnarray}
{\cal A}(k) = 1 + {\frac {{\widetilde{\cal V}}_k} 
{{\widetilde{\cal V}}}}{\left[ {\widetilde{\cal V}}_k + 
i \lambda {\widetilde{\cal I}} 
{\left( 1 + \Omega_g \right)} \right]}, 
\label{Acoeff} 
\end{eqnarray}

\begin{eqnarray}
{\cal B}(k) = {\widetilde{\cal L}} 
{\left \langle {\frac {1} {{\widetilde{\cal F}}}} 
{\frac {\partial S_F} {\partial u}} 
\right \rangle} - \lambda {\widetilde{\cal L}} S_V 
{\left \langle {\frac {1} {{\widetilde{\cal F}}}} 
{\frac {\partial} {\partial u}} {\left( 
{\frac {1} {{\widetilde{\cal F}}^{\ast}}} 
{\frac {\partial f_0} {\partial u}} \right)} 
\right \rangle}, 
\label{Bcoeff} 
\end{eqnarray}

\begin{eqnarray}
{\cal C}(k) = {\widetilde{\cal L}} 
{\left \langle {\frac {1} {{\widetilde{\cal F}}}} 
{\frac {\partial} {\partial u}} {\left( 
{\frac {1} {{\widetilde{\cal F}}}} 
{\frac {\partial f_0} {\partial u}} \right)} 
\right \rangle}, 
\label{Ccoeff} 
\end{eqnarray}

\noindent
and the functional ${\cal G}{\left( [F] \right)}$ of the 
amplitude $F$ can be written as 

\begin{eqnarray}
{\cal G}{\left( [F] \right)} = {\left \langle {\frac {1} 
{{\widetilde{\cal F}}}} {\frac {\partial F} {\partial u}} 
\right \rangle}. 
\label{Couple} 
\end{eqnarray}

Equations (\ref{Amplitude1}) and (\ref{Amplitude2}) comprise 
the system of coupled amplitude equations for the intensity 
of a resonator wave with a wave number $k$ and the slowly 
varying part of the longitudinal distribution function. Note 
that the dependence on the temporal scale $T_1$ (involving 
derivatives with respect to $T_1$) in equations 
(\ref{Amplitude1}) and (\ref{Amplitude2}) through the 
operator ${\widehat{\cal L}}_1$ and the function $f_V$ can 
be eliminated in principle by using the first order equations 
(\ref{Symmetry2}) and (\ref{Symmetry}). As a result one 
obtains a system of coupled second order partial differential 
equations for $F$ and $E$ with respect to the variables 
$T_2$ and $z_1$.

\section{Concluding Remarks}

We have studied the longitudinal dynamics of particles moving 
in an accelerator under the action of a collective force due 
to a resonator voltage. For a sufficiently high beam density 
(relatively large value of the parameter $\lambda$) the 
nonlinear wave coupling, described by the nonlinear term in 
the Vlasov equation becomes important, and has to be taken 
into account. This is manifested in a spatio-temporal 
modulation of the wave amplitudes in unison with the slow 
process of particle redistribution. As a result of this 
wave-particle interaction (coupling between resonator waves 
and particle distribution modes) coherent, self-organized 
patterns can be formed in a wide range of relevant parameters. 

We have analyzed the slow evolution of the amplitude of a 
single resonator wave with an arbitrary wave number $k$ (and 
wave frequency $\Omega(k)$ defined as a solution to the 
dispersion relation). Using the method of multiple scales 
a system of coupled amplitude equations for the resonator 
wave envelope and for the slowly varying part of the 
longitudinal distribution function has been derived. As 
expected, the equation for the resonator wave envelope is a 
generalized cubic Ginzburg--Landau (GCGE) equation. We 
argue that these amplitude equations govern the (relatively) 
slow process of formation of coherent structures and 
establishment of wave-particle equilibrium.

\subsection*{Acknowledgments}

The author wishes to thank Y. Oono and C. Bohn for careful 
reading of the manuscript and for making valuable comments. 

This work was supported by the US Department of Energy, 
Office of Basic Energy Sciences, under contract 
DE-AC03-76SF00515.


\begin{thebibliography}{9}

\bibitem{chao} A.W. Chao, {\it Physics of Collective Beam 
Instabilities in High-Energy Accelerators}, Wiley, New York, 
1993. 

\bibitem{az1} P.L. Colestock, L.K. Spentzouris and S.I. Tzenov, 
In {\it Proc. International Symposium on Near Beam Physics}, 
Fermilab, September 22--24, 1997, FNAL-Conf-98/166, 1998, 
pp. 94--104. 

\bibitem{gerasimov} A. Gerasimov, Phys. Rev. E {\bf 49}, 
(1994), p. 2331. 

\bibitem{az2} S.I. Tzenov and P.L. Colestock, FNAL-Pub-98/258, 
1998. 

\bibitem{az3} S.I. Tzenov, FNAL-Pub-98/275, 1998. 

\bibitem{az4} S.I. Tzenov, In {\it Proc. Workshop on 
Instabilities of High Intensity Hadron Beams in Rings}, 
Upton, New York, June/July 1999, T. Roser and S.Y. Zhang eds., 
AIP Conf. Proc. {\bf 496}, 1999, pp. 351--360.

\bibitem{cary} P.H. Stoltz and J.R. Cary, {\it Physics of 
Plasmas}, {\bf 7}, (2000), p. 231. 

\bibitem{spentzouris} L.K. Spentzouris, Ph.D. Thesis, 
Northwestern University, 1996. 

\bibitem{bruck} H. Bruck, {\it Accelerateurs Circulaires de 
Particules}, Presses Universitaires, Paris, 1966. 

\bibitem{dodd} R.K. Dodd, J.C. Eilbeck, J.D. Gibbon and 
H.C. Morris, {\it Solitons and Nonlinear Wave Equations}, 
Academic Press, London, 1982. 

\bibitem{debnath} Lokenath Debnath, {\it Nonlinear Partial 
Differential Equations for Scientists and Engineers}, 
Birkhauser, Boston, 1997. 

\end{thebibliography}
\end{document}